\documentclass[a4paper, 10pt]{article}

\usepackage[left=2cm,right=2cm,top=2cm,bottom=2cm]{geometry}
\usepackage{graphicx}
\usepackage{amsmath}
\usepackage[utf8]{inputenc}
\usepackage{hyperref}
\usepackage{url}
\usepackage{amssymb}
\usepackage{bm}
\usepackage{amsfonts}
\usepackage{indentfirst}
\usepackage[toc,title]{appendix}
\usepackage[font=small]{caption}
\usepackage{cite}
\usepackage{xcolor}

\usepackage{tikz}

\def\cont{
\tikz[baseline=.1ex]{
\draw (0.4ex,-0.99ex+0.6ex) -- (-0.3ex,0+0.6ex);
\draw[->] (-0.3ex,0+0.6ex) -- (0.7ex,0.99ex+0.6ex);
}
}

\def\contSecond{
\tikz[baseline=.1ex]{
\draw (-0.4ex,-0.99ex+0.6ex) -- (0.3ex,0+0.6ex);
\draw[->] (0.3ex,0+0.6ex) -- (-0.7ex,0.99ex+0.6ex);
}
}

\def\contMinusFirst{
\tikz[baseline=.1ex]{
\draw (0.4ex,0.99ex+0.6ex) --(-0.3ex,0+0.6ex) ; %%% upper-left to middle
\draw [->] (-0.3ex,0+0.6ex) -- (0.7ex,-0.99ex+0.6ex);   %%% middle to lower-right
}
}

\newcommand{\tr}{\textup{Tr}}

\newcommand{\<}{\left<}
\renewcommand{\>}{\right>}

\newcommand{\cP}{{\cal P}}

\newcommand{\ket}[1]{\left| #1 \>}
\newcommand{\bra}[1]{\< #1 \right|}
\newcommand{\braket}[2]{\< #1 | #2 \>}

\newcommand{\He}{\textup{He}}

\newcommand{\tR}{\tilde{R}}
\newcommand{\tP}{\tilde{P}}

\newcommand{\tS}{\tilde{S}}
\newcommand{\Ai}{\mathrm{Ai}}
\newcommand{\Bi}{\mathrm{Bi}}
\newcommand{\Hi}{\mathrm{Hi}}

\usepackage{amsthm}

\newtheorem{theorem}{Theorem}[section]
\newtheorem{corollary}[theorem]{Corollary}
\newtheorem{proposition}[theorem]{Proposition}

\newtheorem{lemma}[theorem]{Lemma}

\theoremstyle{definition}

\newtheorem{remark}[theorem]{Remark}

\begin{document}

\title{Condition numbers for real eigenvalues of real elliptic ensemble: weak non-normality at the edge}
\author{Wojciech Tarnowski\footnote{wojciech.tarnowski@doctoral.uj.edu.pl}
\\[0.5ex]
 {\small Marian Smoluchowski Institute of Physics,} \\
\small{ Jagiellonian University, S. \L ojasiewicza 11, PL 30-348 Krak\'ow, Poland}
}

\date{\today}
\maketitle

\begin{abstract}
Sensitivity of an eigenvalue $\lambda_i$ to the perturbation of matrix elements is controlled by the eigenvalue condition number defined as $\kappa_i = \sqrt{\braket{L_i}{L_i}\braket{R_i}{R_i}}$, where $\bra{L_i}$ and $\ket{R_i}$ are left and right eigenvectors to the eigenvalue $\lambda_i$. In Random Matrix Theory the squared eigenvalue condition number is also known as the eigenvector self-overlap. In this work we calculate the asymptotics of the joint probability density function of the real eigenvalue and the square of the corresponding eigenvalue condition number for the real elliptic ensemble in the double scaling regime of almost Hermiticity and close to the edge of the spectrum. As a byproduct, we also calculate the one-parameter deformation of the Scorer's function. 
\end{abstract}

\section{Introduction}

Statistical properties of eigenvalues of asymmetric real matrices have been extensively studied through the lenses of Random Matrix Theory (RMT) over the past 60 years. Starting from the pioneering work of Ginibre~\cite{Ginibre}, Gaussian ensembles were analyzed~\cite{Sommers,Edelman1,Edelman2}, followed by the discovery of the Pfaffian structure of eigenvalue densities~\cite{AkemannKanzieper,BorodinSinclair,ForresterNagao,BorodinSinclair2}. New solvable ensembles were found~\cite{ForresterElliptic,TruncatedOrthogonal,Spherical,InducedGinibre} and development of new RMT tools allowed for the study of products of random matrices~\cite{Products1,Products2,Products3,Products4}. Due to the universality of their global densities as well as universal microscopic correlations, random matrices are widely used in complex systems~\cite{May,MayNonlinear, MayBouchaud} and quantum chaos~\cite{QChaos1,QChaos2}.

Much less, however, is known about the properties of their eigenvectors. Early studies followed the spirit of symmetric RMT and focused mostly on the localization properties~\cite{Localization1}. Dropping the symmetry or, in general, normality (a matrix is normal if it commutes with its transpose), opens a new dimension for eigenvectors, because for each eigenvalue $\lambda_i$ there is not one, but two eigenvectors -- left $\bra{L_i}$ and right $\ket{R_i}$ -- satisfying their own eigenproblems: $X\ket{R_i} = \lambda_i \ket{R_i}$ and $\bra{L_i}X = \bra{L_i}\lambda_i$. In contrast to the normal case, right (left) eigenvectors are not orthogonal to each other, $\braket{R_i}{R_j}\neq \delta_{ij}\neq\braket{L_i}{L_j}$. Instead, left and right eigenvectors are normalized to $\braket{L_i}{R_j}=\delta_{ij}$, forming a biorthogonal set. 

Traces of eigenvector non-orthogonality are observed in open quantum  chaotic systems through the decay laws~\cite{SavinSokolov}, excess noise in open laser resonators~\cite{RandomLasing1,RandomLasing2}, resonance width shifts\cite{WidthShifts,WidthShiftsExp}, and in the shape of reflected power profiles~\cite{CPAFyodorov}. In dynamical systems early-time transients  are driven by the eigenvectors~\cite{GrelaTransient}. Transient amplification of noise was proposed as a mechanism behind the formation of Turing patterns~\cite{Turing1,Turing2,Turing3}. In theoretical neuroscience transient dynamics was proposed as a mechanism for amplification of weak neuronal signals~\cite{Neuro1, Neuro2, Neuro3} with some biologically inspired models exhibiting strong eigenvector non-orthogonality~\cite{NECO,WTtransient}. Furthermore, non-orthogonal eigenvectors play a role in Dysonian dynamics in non-Hermitian matrices~\cite{Diffusion1, Diffusion2, DysonFull1, BourgadeDubach, DysonFull2} and generalizations of fluctuation-dissipation relations to non-equilibrium systems~\cite{FDT1,FDT2}. Recently, they appeared in the context of localization transition in non-Hermitian systems~\cite{LocalizationUeda, LocalizationGosh} and the eigenstate thermalization hypothesis~\cite{NonhermitianETH}.

The study of eigenvector non-normality in RMT was initiated by Chalker and Mehlig~\cite{ChalkerMehlig,ChalkerMehlig2} who introduced the matrix of overlaps $O_{ij}=\braket{L_i}{L_j}\braket{R_j}{R_i}$, though this matrix was known before in nuclear physics as the Bell-Steinberger matrix~\cite{BellSteinberger}. Its diagonal elements, often referred to as \textit{self-overlaps}, are the Petermann factors in random lasing~\cite{Petermann} and squares of the \textit{eigenvalue condition numbers}~\cite{Wilkinson,SingleRing} in numerical analysis. Since the full distribution of overlaps is difficult to study, Chalker and Mehlig managed to calculate it only for $N=2$ and turned the attention to their mean values by introducing the correlation functions
\begin{equation}
O_1(z) = \< \frac{1}{N^2}\sum_{k=1}^{N} O_{kk}\delta^{(2)}(z-\lambda_k) \>,\qquad O_2(z,w) = \<\frac{1}{N} \sum_{\substack{k,l=1 \\ k\neq l}}^{N} O_{kl} \delta^{(2)}(z-\lambda_k)\delta^{(2)}(w-\lambda_l)\>.
\end{equation}
These objects are more tractable and became accessible with the use of large $N$ diagrammatics~\cite{NowakJanikEigenvectors,MehligSanter,NowakTarnowskiProbing}, supersymmetry~\cite{FyodorovMehlig,RandomLasing1,RandomLasing2,FrahmSchomerus} and explicit calculations at finite size~\cite{WaltersStarr, BurdaVivo, AkemannForster, AkemannZaboronski, WurfelCrumpton, NodaDeterminantal, NodaDeterminantal2,FyodorovWurfel}. Recently, they were also extended to higher-order correlation functions~\cite{CrawfordRosenthal}.

 Recently, Dubach and Bourgade~\cite{BourgadeDubach} and Fyodorov~\cite{FyodorovBiorthogonal} calculated the joint probability density function (jpdf) of the eigenvalues and corresponding eigenvector self-overlap in the complex Ginibre ensemble. Fyodorov obtained also the jpdf 
\begin{equation}
\cP_N(z,t) = \< \sum_{k} \delta(t + 1- O_{kk})\delta(z-\lambda_k) \> \label{eq:jpdf}
\end{equation}
for the real eigenvalues in the real Ginibre, where the sum is performed over real eigenvalues and jpdf is normalized to the total number of real eigenvalues. Note that since $O_{ii}\geq 1$, it is more convenient to consider a shifted overlap $t=O_{ii}-1$. Subsequent results were also obtained for various other ensembles~\cite{Dubach2,Dubach3,FyodorovTarnowski,Dubach4} and the eigenvector non-orthogonality was approached from other directions as well~\cite{Benaych}.

%suggest that the large $N$ limit of the jpdf is universal and takes the form of an inverse gamma distribution $c^\beta(z) \rho(z)t^{-\beta-1} \exp(-c(z)/t)$ with the scale parameter $c(z)=\frac{O_1(z)}{\rho(z)}$. Here $\rho(z)$ is the eigenvalue density and $\beta=1$ for real eigenvalues of real matrices, $\beta=4$ for symplectic ensembles and $\beta=2$ for complex ensembles and for complex eigenvalues of real ensembles~\cite{}. The scale parameter is the conditional expectation of $O_{ii}$ for $\beta=2,4$, however for real eigenvalues ($\beta=1$) the expectation does not exist due to the heavy tail. Nevertheless, the relation holds if one uses $O_1(z)$ calculated for complex eigenvalues and extends it analytically to the real line.

For symmetric (in general normal) matrices the overlap matrix reduces to the identity matrix. It is therefore tempting to study eigenvector non-orthogonality at the transition between symmetric and asymmetric matrices. A natural model for such an analysis is the real elliptic ensemble. Elements of such matrices are Gaussian and distributed according to the measure
\begin{equation}
P(X)dX = C_N^{-1} \exp\left( - \frac{1}{2(1-\tau^2)} \tr (XX^T - \tau  X^2)\right) dX, \label{eq:EllipticPDF}
\end{equation}
where $dX = \prod_{ij=1}^{N} dx_{ij}$ is the flat Lebesgue measure over its elements. The normalization constant reads $C_N=(2\pi)^{N^2/2}(1+\tau)^{N/2}(1-\tau^2)^{N(N-1)/4}$. The parameter $\tau \in [0,1]$ controls the correlation between elements on the opposite sides of the diagonal and provides continuous interpolation between the real Ginibre ensemble for $\tau=0$ and the Gaussian Orthogonal Ensemble (GOE)  for $\tau=1$. In the limit $N\to \infty$ for any fixed $0\leq \tau <1$ the eigenvalue statistics fall into the bulk universality class of non-symmetric matrices. However, after tuning the rate of approaching symmetry so that the product $N(1-\tau)$ is kept fixed  one finds a new regime interpolating between GOE sine-kernel universality and non-Hermitian bulk universality. This regime was first studied for the complex elliptic ensemble and was dubbed \textit{weak non-Hermiticity}~\cite{WeakNonH1,WeakNonH2,WeakNonH3} (see~\cite{ForresterElliptic} for the results in the real case).

Recently, an analogous regime was found at the level of jpdf~\eqref{eq:jpdf} for the real eigenvalues of the real elliptic ensemble. The self-overlap transitions from $O_{ii}=1$ at $\tau=1$ to $O_{ii}\sim N$ for fixed $\tau<1$. At the transition regime, called \textit{weak non-normality}, the overlap is of order 1 with a non-trivial heavy-tailed density. The heavy-tailedness was found to be the most robust feature of self-overlap distribution, as it appears also for finite $N$~\cite{ChalkerMehlig} and rank-1 perturbations of Hermitian matrices~\cite{CPAFyodorov}.

The jpdf for the elliptic ensemble can be written in terms of the joint density $P_N^{\tau}(z,q)$ of an eigenvalue $z$ and rescaled and shifted overlap $q=(1-\tau)  (O_{ii}-1)$. The jpdf~\eqref{eq:jpdf} is recovered via $\cP_N(z,t) = (1-\tau)^{-1} \cP_N^{\tau}(z,\frac{t}{1-\tau})$ and $\cP_N^{\tau}$ reads~\cite[Theorem 2.1]{FyodorovTarnowski}
\begin{equation}
\cP_N^{\tau}(z,q) = \frac{1}{2(1+\tau)\sqrt{2\pi}} \frac{ e^{-\frac{z^2}{2(1+\tau)}\left(1+\frac{q}{1+q}\right)}}{\sqrt{q(q+1)}} \left(\frac{q}{q+1+\tau}\right)^{\frac{N}{2}-1} Q_N(z,q,\tau), \label{eq:PEllipticFull}
\end{equation}
where
\begin{multline}
Q_N(z,q,\tau) = \frac{(1+\tau-2z^2)\tP_{N-1}(z) + z\tR_{N-1}(z) + \tau z \tR_{N-2}(z)}{1+q} + \frac{z^2\tP_{N-1}(z)}{(1+q)^2} + \frac{\tau^2(1+\tau)^2 N\tP_{N-2}(z)}{(1+\tau+q)^2}  
\\
+\frac{(1+\tau)(1-\tau^2)\tS_{N-2}(z)}{1+\tau+q} - \frac{\tau(1+\tau)z\tR_{N-2}(z)}{(1+q)(1+\tau+q)} \label{eq:QN}
\end{multline}
and
\begin{eqnarray}
\tP_N(z) & = & \sum_{k=0}^{N-1} \frac{1}{k!}((k+1)p_k^2(z)-kp_{k+1}(z)p_{k-1}(z)), \label{eq:P}
\\
\tR_N(z) & = & \sum_{k=0}^{N-1} \frac{1}{k!}((k+2)p_{k+1}(z)p_k(z)-kp_{k+2}(z)p_{k-1}(z)), \label{eq:R}
\\
\tS_N(z) & = & \sum_{k=0}^{N-1} \frac{N-k}{k!}( (k+1)p_k^2(z)-kp_{k+1}(z)p_{k-1}(z)). \label{eq:S}
\end{eqnarray}
Here $p_k(z) = \tau^{k/2}\He_k\left(\frac{z}{\sqrt{\tau}}\right) = \left(\frac{\tau}{2}\right)^{k/2}\mathrm{H}_k\left(\frac{z}{\sqrt{2\tau}}\right)$ are the Hermite polynomials with the leading term $z^k$.

\begin{remark}
The presented form slightly differs from the one in~\cite{FyodorovTarnowski}, hence we used tilded variables to distinguish from the original (non-tilded) notation. They are related as follows 
\begin{displaymath}
P_N(z) = N!\tP_{N+1}(z),\qquad
2R_N(z) = N!\tR_{N+1}(z),\qquad
NP_{N-1}(z) - T_{N-1}(z) = (N-1)! \tS_N(z).
\end{displaymath}
\end{remark}

\section{Statement and discussion of the main results}

The elliptic ensemble renders yet another weak non-Hermiticity regime.  At $\tau=1$ near the edge of the spectrum eigenvalue statistics are described by the Airy kernel, while at strong non-Hermiticity the kernel falls into the error function universality class. The interpolation between those two was found much later after the bulk weak non-Hermiticity~\cite{Bender,AkemannPhillips,AkemannPhillips2}. The edge behavior is particularly interesting, since the rightmost (i.e. the one with the largest real part) eigenvalue plays an essential role in the stability of linear systems.
 It is therefore natural to ask how stable  the eigenvalues are at the edge of the spectrum. An insight into this problem can be obtained by analyzing the edge behavior of jpdf~\eqref{eq:jpdf}. Recent work~\cite{FyodorovTarnowski} studied the limiting distribution of \eqref{eq:PEllipticFull} at strong non-normality. 
The main result of this work is complementing that calculation with the double scaling regime in which the spectral edge is probed on the scale $N^{-1/6}$ and departure from symmetry on the scale $N^{-1/3}$.

Before we present the result, let us define the deformed Airy function, which has appeared in the study of weak non-Hermiticity at the edge~\cite{AkemannPhillips,ByunLee}, as
\begin{equation}
\Ai_b(\zeta) := \frac{1}{2\pi i} \int\limits_{\cont}  e^{\frac{u^3}{3}+\frac{b^2u^2}{2}-u\zeta} du = e^{\frac{b^2\zeta}{2}+\frac{b^6}{12}} \Ai\left(\zeta + \frac{b^4}{4}\right). \label{eq:AiryDeformed}
\end{equation}
The integration contour denoted schematically as $\cont$ starts at $\infty e^{-i\pi/3}$ and ends at $\infty e^{+i\pi/3}$, which is a standard choice for the Airy function of a complex argument. The second equality in \eqref{eq:AiryDeformed} is obtained by shifting the integration variable $u\to u - b^2/2$. For $b=0$ the function $\Ai_b$ reduces to the Airy function.

\begin{theorem} \label{th:Main}
Let $\cP_N(z,t)$ be defined as in \eqref{eq:jpdf}, where the ensemble average $\<...\>$ is taken with respect to the real elliptic ensemble given by \eqref{eq:EllipticPDF}. Let $\cP^{w.e.}(\zeta,t)$ be the limit of $N^{-1/6}\cP_N(z=\sqrt{N}(1+\tau)+\zeta N^{-1/6}, t)$ when $N\to\infty$ and $\tau \to 1$ such that $(1-\tau)N^{1/3} = b^2$ remains fixed. The limit reads
\begin{equation}
\cP^{w.e.}(\zeta,t) = \frac{b^2}{t^2}  \left[ T_0(\zeta) + \frac{b^2 T_1(\zeta)}{t}+\frac{b^4 T_2(\zeta)}{t^2}+\frac{b^6 T_3(\zeta)}{t^3} \right]\exp\left(\frac{b^2\zeta}{t} - \frac{b^6}{2t^2} - \frac{b^6}{3t^3}\right) \label{eq:Main}
\end{equation}
where
\begin{eqnarray}
T_3(\zeta) & = &  \int_{\zeta}^{\infty} \left[ \Ai_b'^2(p) -  \Ai_b(p)\Ai_b''(p)\right] dp \label{eq:T3}
\\
T_2(\zeta) & = & b^2 T_3 + \int_{\zeta}^{\infty} \Ai_b^2(p) dp
\\
T_1(\zeta) & = &  - \zeta T_3 + \frac{1}{2} \Ai_b^2(\zeta)  + b^2 \int_{\zeta}^{\infty} \Ai_b^2(p) dp
\\
T_0(\zeta) & = & - T_3 +\frac{1}{2}b^2 \Ai_b^2(\zeta) - \frac{1}{2} \Ai_b(\zeta)\Ai_b'(\zeta)  - \zeta \int_{\zeta}^{\infty} \Ai_b^2(p)dp. \label{eq:T0}
\end{eqnarray}
\end{theorem} 

In analogy to $\cP_{N}(z,t)$ taking a simpler form when $t$ is rescaled by the departure from normality, rescaling $t \to t/ b^{2}$ slightly simplifies \eqref{eq:Main}.
The jpdf~\eqref{eq:jpdf} is a 2-point probability density function, which after integrating out the overlap component yields the edge density at weak non-Hermiticity~\cite{AkemannPhillips2,ByunLee}.
\begin{corollary}\label{th:Integration}
Let $\cP^{w.e.}$ be defined as in Theorem \ref{th:Main}. The following property holds
\begin{equation}
\rho_b(\zeta) := \int_0^{\infty}\cP^{w.e.}(\zeta,t) dt = \int_{\zeta}^{\infty} \Ai_b^2(p)dp + \frac{1}{2}\Ai_b(\zeta)\left(1-\int_{\zeta}^{\infty} \Ai_b(p)dp \right). \label{eq:density}
\end{equation}
\end{corollary}
It is clearly visible from \eqref{eq:Main} that  $\lim_{b\to 0} \cP^{w.e.}(\zeta,t)=0$ for any $t>0$, however $\cP^{w.e.}$ is singular at $t=0$. Corollary~\ref{th:Integration} shows that this singularity is integrable, and thus $\cP^{w.e.}(\zeta,t)$ tends to $\rho_0(\zeta)\delta(t)$ as $b\to 0$, in agreement with the eigenvector orthogonality in the symmetric limit.

The first step in the proof of Corollary~\ref{th:Integration} is the change of integration variable $u=\frac{b^2}{t}$. This results in a non-trivial integral, which can be considered as one-parameter deformation of the Scorer's function~\cite{Scorer} (see also~\cite[chapter 2.3]{AiryBook}). The evaluation of this integral is a result that deserves its own interest.

\begin{lemma}\label{th:Scorers}
 Let $b\in\mathbb{R}$ and  define  $\Hi_b(\zeta):=\pi^{-1}\int_0^{\infty} \exp\left(\zeta u - \frac{1}{2}b^2 u^2 - \frac{1}{3} u^3\right)du$. The following result holds
\begin{equation}
\Hi_b(\zeta) =  \Bi_{ib}(\zeta)\int_{-\infty}^{\zeta} \Ai_b(t) dt -\Ai_{ib}(\zeta)\int_{-\infty}^{\zeta} \Bi_b(t) dt, \label{eq:Scorer}
\end{equation}
where $\Bi_b(\zeta) := e^{\frac{b^2\zeta}{2}+\frac{b^6}{12}} \Bi\left(\zeta + \frac{b^4}{4}\right)$ and $\Bi$ is the second solution of the Airy equation.
\end{lemma}

Interestingly, the scaling in Theorem~\ref{th:Main} does not involve the variable $t$. This means that the self-overlap fluctuates around $1$ on the scale $\mathcal{O}(1)$. Exactly the same scale is found at weak non-normality in the bulk~\cite{FyodorovTarnowski}. This stays in contrast with the strong non-normality, where the self-overlap in the bulk grows linearly with the matrix size~\cite{ChalkerMehlig,BourgadeDubach, FyodorovBiorthogonal}, while at the edge its growth is slower ($\sim\sqrt{N}$)~\cite{WaltersStarr,FyodorovBiorthogonal,FyodorovTarnowski}. However,  the two regimes of weak non-normality are achieved at different symmetry-breaking scales. Weak non-normality in the bulk requires parameterization $1-\tau = a^2/2N $, while at the edge the departure from symmetry is larger, i.e. $1-\tau =b^2 N^{-1/3}$. The transition between these two regimes can be probed by letting the parameter $b$ tend to $0$ at the speed $N^{-1/3}$. More precisely, we parameterize $b = \frac{a}{\nu \sqrt{2}}$. At the same time, $\zeta$ which probed the scale of $N^{-1/6}$ needs to be rescaled to probe the macroscopic scale $N^{1/2}$. This is achieved by parameterizing $\zeta= -\nu^2w$.

\begin{corollary} \label{th:BulkWeak}
Let $\cP^{w.e.}$ be defined as in Theorem \ref{th:Main}, let $b = \frac{a}{\nu\sqrt{2}}$ and denote $A=a^2w$ for simplicity. The following limit holds
\begin{equation}
\lim_{\nu\to \infty} \nu^{-1}\cP^{w.e.}(-\nu^2 w, t) = \frac{A \sqrt{w}}{2\pi t^2} e^{-\frac{A}{2t}}\left[ \left( \frac{2}{A} - \frac{1}{t}\right) e^{-\frac{A}{2}} + \left(1+\frac{1}{t} - \frac{2}{A}\right) \int_0^{1} e^{-\frac{As^2}{2}} ds \right]. \label{eq:BulkWeak}
\end{equation}
\end{corollary}

Note here that the Jacobian brings a factor of $\nu^2$, the absorption of which requires rescaling by $\nu^{-3} \sim N^{-1}$. This is a consequence of the normalization of jpdf to the total number of real eigenvalues and the fact that in the weak non-Hermiticity regime almost all eigenvalues are real. The above result can also be obtained from \cite[Remark 2.7]{FyodorovTarnowski} by edge parameterization $z=2-w$.

\begin{figure}[h]
\begin{center}
\includegraphics[width=0.49\textwidth]{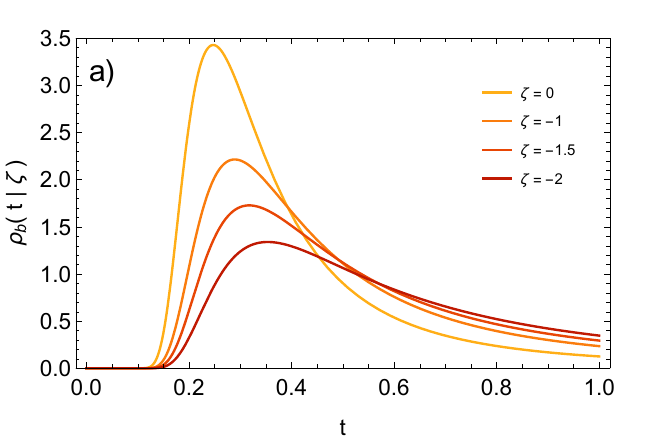}
\includegraphics[width=0.49\textwidth]{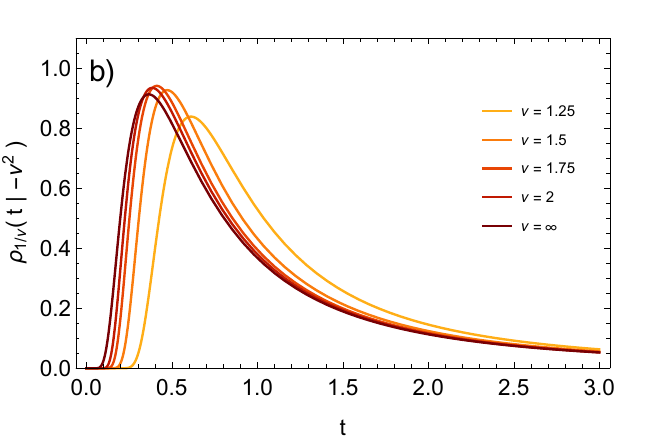}
\end{center}
\caption{a) Density of overlaps conditioned (see \eqref{eq:CondDens}) at various values of the corresponding eigenvalue with $b=0.6$. As $\zeta$ moves deeper into the bulk, the density maxima move to the larger values, as well as the right tail of the distribution lifts up, a sign of increasing eigenvector non-orthogonality. b) Conditional density of the overlap transitioning from edge to bulk weak non-normality, corresponding to scaling $\zeta=-\nu^2$ and $b = 1/\nu $, as in Corollary~\ref{th:BulkWeak}, where we set $a=\sqrt{2}$ and $w=1$. The limiting ($\nu=\infty$) curve is plotted with the use of jpdf at bulk weak non-normality~\eqref{eq:BulkWeak} and eigenvalue density obtained after integrating out the $t$ variable.
\label{fig:Plots}}
\end{figure}

Despite the fact that the self-overlap stays $\mathcal{O}(1)$ in both regimes of edge and bulk weak non-normality, the eigenvector non-orthogonality increases as one moves deeper into the bulk. Fig.~\ref{fig:Plots} shows that if $b$ is kept constant and $\zeta$ is moved to the bulk, the peak of the overlap distribution moves to the right as well as the right tail gets heavier. To account for the fact that $\cP(z,t)$ encodes also the density of eigenvalues, which changes with $\zeta$, we focused on the density of the overlap, conditioned on the eigenvalue, i.e.
\begin{equation}
\rho_b(t|\zeta) = \frac{\cP^{w.e.}(\zeta,t)}{\rho_b(\zeta)}, \label{eq:CondDens}
\end{equation}
where $\rho_b(\zeta)$ is given by \eqref{eq:density}. Despite the fact that the matrix is very close to normal, eigenvectors are unlikely to be orthogonal. As shown in Fig.~\ref{fig:Plots}, the conditional probability density is almost zero for small values of $t$. More precisely, it is exponentially suppressed by the factor $\exp(-b^6/t^3)$.

\begin{figure}
\includegraphics[width=0.5\textwidth]{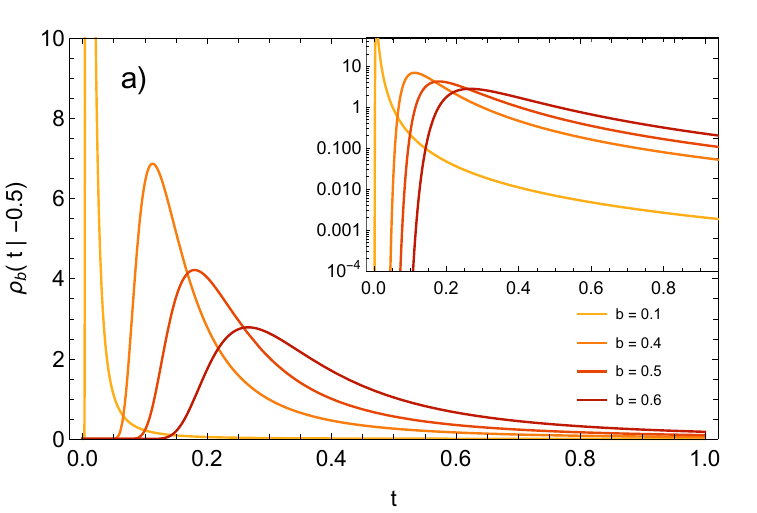}
\includegraphics[width=0.5\textwidth]{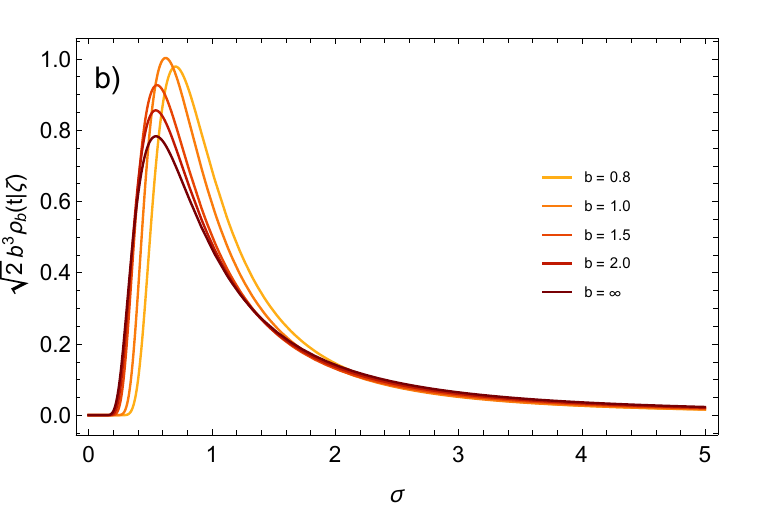}
\caption{Transition from weak to strong non-normality. a) Density of overlaps conditioned at $\zeta=-0.5$ for various values of $b$, showing the evolution of the density as non-normality increases. Inset presents the same densities on the logarithmic $y$ axis.
b)  Conditional density of the overlap with the increasing non-normality in the rescaled variables $\zeta=\sqrt{2}b\delta$ (where we set $\delta=-0.5$) and $t=\sqrt{2}b^3\sigma$, corresponding to the Corollary~\ref{th:StrongNonnormality}. The density was multiplied by $\sqrt{2}b^3$ to account for the Jacobian. The limiting ($b=\infty$) curve was plotted with the use of jpdf at strong non-normality~\eqref{eq:StrongNonnormality} and eigenvalue density obtained after integrating out the $t$ variable.
\label{fig:PlotsTransition}
}
\end{figure}

With the increasing non-normality, the distribution of the overlap moves towards larger values of $t$ (see Fig.~\ref{fig:PlotsTransition}), up to the strong non-normality regime. This is obtained from~\eqref{eq:PEllipticFull} by parameterizing $z=\sqrt{N}(1+\tau) + \delta\sqrt{1-\tau^2}$ and $t = \sigma \sqrt{N(1+\tau)/(1-\tau)}$~\cite{FyodorovTarnowski}. The limiting function can be recovered 
from Theorem~\ref{th:Main} by letting $b\to \infty$. The appropriate rescaling of $t$ and $\sigma$ can be deduced from the fact that in order to achieve $1-\tau = \mathcal{O}(1)$, the parameter $b$ has to be of order $N^{1/6}$, which leads to the parameterization $\zeta=\sqrt{2}b\delta$ and $t=\sqrt{2} b^3\sigma$.

\begin{corollary} \label{th:StrongNonnormality}
Let $\cP^{w.e.}$ be defined as in Theorem \ref{th:Main}. It admits the following limit
\begin{equation}
\lim_{b\to\infty} 2b^4\cP^{w.e.}(\sqrt{2}b\delta,\sqrt{2}b^3\sigma) = \frac{1}{4\pi \sigma^2} e^{\frac{\delta}{\sigma} - \frac{1}{4\sigma^2}}\left( e^{-2\delta^2} + \left(\frac{1}{\sigma}-2\delta\right) \int\limits_{2\delta}^{\infty} e^{-\frac{p^2}{2}} dp \right).
\label{eq:StrongNonnormality}
\end{equation}
\end{corollary}
This formula stays in full agreement with known results~\cite{FyodorovBiorthogonal,FyodorovTarnowski}. The prefactor $2b^4$ is the Jacobian of the reparameterization.

\begin{remark} \label{remark}
The transition from symmetry to strong non-normality can be interpreted as a diffusion-like process.  The deformed Airy function obeys diffusion equation in which $\eta = b^2/2$ plays the role of time. More precisely, the function $\varphi(\zeta,\eta) = \exp(\zeta \eta + \frac{2}{3}\eta^3)\Ai(\zeta + \eta^2)$ satisfies $\partial_{\eta}\varphi = \partial_{\zeta\zeta}\varphi$. This also explains the appearance of Gaussians in Corollary~\ref{th:StrongNonnormality}.
\end{remark}

The remaining part of the paper is devoted to the proofs of the main results and is organized as follows. Section \ref{sec:int_rep} introduces  integral representation useful for the asymptotic analysis of \eqref{eq:PEllipticFull}. The proof of Theorem~\ref{th:Main}, which essentially relies on the saddle point analysis of~\eqref{eq:QN} is presented in section \ref{sec:saddle_point}. Proofs of Corollaries~\ref{th:BulkWeak} and~\ref{th:StrongNonnormality}  are presented in sections~\ref{sec:BulkWeak} and~\ref{sec:strong_nonnorm}, respectively, while section~\ref{sec:integrating_out} is devoted to the proof of Corollary~\ref{th:Integration} and Lemma~\ref{th:Scorers}.

\section{Proofs of the results}
\subsection{Integral representations} \label{sec:int_rep}

\begin{figure}
\begin{center}
\includegraphics[width=\textwidth]{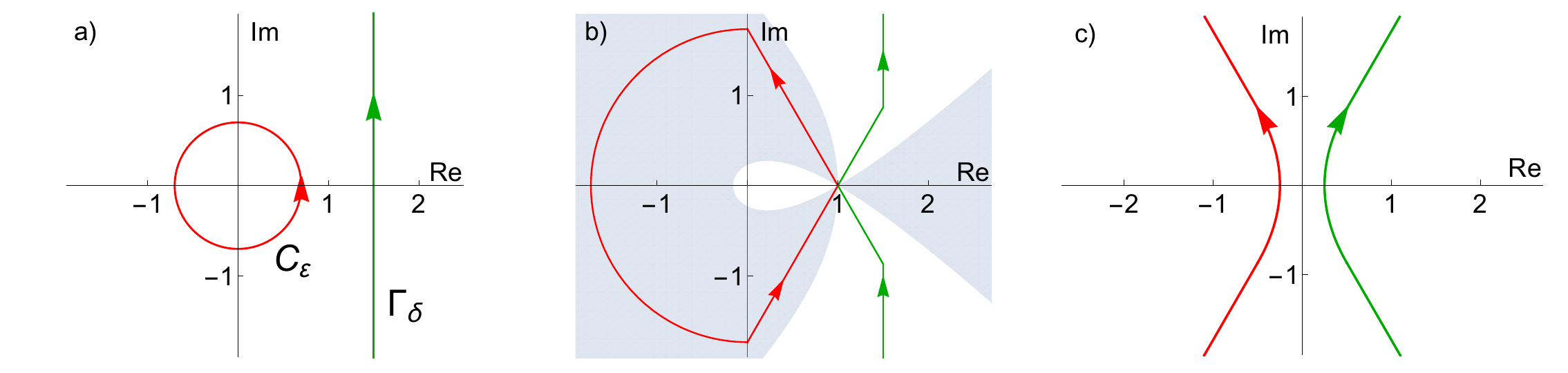}
\end{center}
\caption[aaa]{a) Integration contours $C_{\varepsilon}$ and $\Gamma_{\delta}$ in Lemma~\ref{th:IntegralRepresentation}. b) Contours $C_{\varepsilon}$ and $\Gamma_{\delta}$ are deformed to collect the contribution to the integral from the vicinity of the saddle point at $+1$. Shaded area represents the region where the real part of $\tilde{f}(z) = f(z) + 3/2$ is positive. This means that the integrand in the red contour is exponentially suppressed everywhere except the vicinity of the stationary point. Analogously, in the unshaded region $\textup{Re} \tilde{f} <0$ and the integrand over green contour is exponentially suppressed.  c) With the parameterization $s=1+uN^{-1/3}$ and $w=1+vN^{-1/3}$ we probe the direct vicinity of the stationary point, where contours are approximated by two incoming rays with slopes $e^{-2i\pi/3}$ and $e^{-\pi i/3}$, and two outgoing rays extending to infinity with slopes $e^{2i\pi/3}$ and $e^{i\pi/3}$. Throughout the text the contours are denoted as $\contSecond$ and $\cont$.
\label{fig:contours}
}
\end{figure}

\begin{lemma} \label{th:IntegralRepresentation}
The functions $\tP_N$, $\tR_N$ and $\tS_N$ defined in equations \eqref{eq:P} - \eqref{eq:S} admit the integral representations
\begin{eqnarray}
\tP_N(z) &  = & - \frac{\tau^{-1/2}}{(2\pi)^{3/2}}\int_{\Gamma_{\delta}} ds \oint_{C_{\varepsilon}} dw \frac{s}{s-w}\left(\frac{z-s}{\tau}- w\right) \left(\frac{s}{w}\right)^N e^{\frac{(s-z)^2}{2\tau} - \frac{\tau w^2}{2} + wz}, \label{eq:Prepr}
\\
\tR_N(z) & = & - \frac{\tau^{-1/2}}{(2\pi)^{3/2}} \int_{\Gamma_{\delta}} ds \oint_{C_{\varepsilon}} dw \frac{s}{s-w}\left(\frac{(z-s)^2}{\tau^2}+\frac{1}{\tau}-w^2\right) \left(\frac{s}{w}\right)^{N+1} e^{\frac{(s-z)^2}{2\tau} - \frac{\tau w^2}{2} + wz}, \label{eq:Rrepr}
\\
\tS_N(z) & = & - \frac{\tau^{-1/2}}{(2\pi)^{3/2}} \int_{\Gamma_{\delta}} ds \oint_{C_{\varepsilon}} dw \frac{s^2}{(s-w)^2}\left(\frac{z-s}{\tau}- w\right) \left(\frac{s}{w}\right)^N e^{\frac{(s-z)^2}{2\tau} - \frac{\tau w^2}{2} + wz}, \label{eq:Srepr}
\end{eqnarray}
where the integration contours are parameterized as $\Gamma_{\delta} := \{ is + \delta: s\in \mathbb{R}\}$ and 
$C_{\varepsilon} = \{\varepsilon e^{i\theta}: \theta \in [0,2\pi) \}$ with $\delta > \varepsilon$, see Fig.~\ref{fig:contours}a.
\end{lemma}

\textit{Proof.} We first notice that $\frac{d}{dx}p_k(x) = k p_{k-1}(x)$, which allows to represent $\tP_N$, $\tR_N$ and $\tS_N$ as derivatives of simpler objects as follows
\begin{eqnarray}
\tP_N(z) & = & \lim_{x,y\to z} (\partial_x - \partial_y) U_{N}(x,y), \label{eq:PfromU}\\
\tR_N(z) & = & \lim_{x,y\to z} (\partial^2_x - \partial^2_y) U_{N+1}(x,y), \label{eq:RfromU} \\
\tS_N(z) & = & \lim_{x,y\to z} (\partial_x - \partial_y) V_{N}(x,y). \label{eq:SfromV}
\end{eqnarray}
The axuliary objects are defined as
\begin{equation}
U_N(x,y) = \sum_{k=0}^{N-1} \frac{1}{k!} p_{k+1}(x) p_k(y), \qquad \quad
V_N(x,y) =  \sum_{k=0}^{N-1} \frac{N-k}{k!} p_{k+1}(x) p_k(y). 
\end{equation}
We use two standard integral representations of the Hermite polynomials
\begin{equation}
\mathrm{H}_{k} (x) = \frac{(2i)^k}{\sqrt{\pi}} \int_{\mathbb{R}} s^k e^{-(s+ix)^2}ds,
 \qquad \mathrm{H}_k(y)= \frac{k!}{2\pi i} \oint_{C_{\varepsilon}} \frac{dw}{w^{k+1}} e^{-w^2+2wy}.
\end{equation}
In the first representation the change of variables $s \to -i s$ rotates the contour by $\frac{\pi}{2}$ counterclockwise. Since the integrand is analytic, the contour can be further shifted to $\Gamma_{\delta}$. The choice $\delta > \varepsilon $ ensures that the contours do not intersect.
 The use of $p_k(x) = \left(\frac{\tau}{2}\right)^{k/2} \mathrm{H}_k (\frac{z}{\sqrt{2\tau}})$ leads us to
\begin{equation}
p_k(x) = \frac{1}{\sqrt{2\pi \tau} i} \int_{\Gamma_{\delta}} s^k e^{\frac{1}{2\tau} (s-x)^2} ds,
\qquad \quad 
p_k(y) = \frac{k!}{2\pi i} \oint_{C_{\varepsilon}} \frac{dw}{w^{k+1}} e^{-\frac{\tau w^2}{2} + wy}.
\end{equation}
Therefore, $U_N$ can be represented as 
\begin{equation}
U_N(x,y) = \frac{-1}{(2\pi)^{3/2} \sqrt{\tau} } \int_{\Gamma_{\delta}} ds \oint_{C_{\varepsilon}} dw\, e^{\frac{(x-s)^2}{2\tau}} e^{  - \frac{\tau w^2}{2}+wy} \sum_{k=0}^{N-1} \frac{s^{k+1}}{w^{k+1}}
\end{equation}
The sum is evaluated to
\begin{displaymath}
s\frac{\left(\frac{s}{w}\right)^N-1}{s-w},
\end{displaymath}
however, the term $\frac{s}{s-w}$ does not have a pole at $w=0$, therefore the integral over $w$ will yield 0 for this part. Performing analogous computation for $V_N$, one encounters 
\begin{equation}
\sum_{k=0}^{N-1} (N-k)\frac{s^{k+1}}{w^{k+1}} = \frac{Ns}{w-s} + s^2 \frac{ (\frac{s}{w})^{N}-1}{(s-w)^2}.
\end{equation}
Similarly, only the term $\frac{s^2}{(s-w)^2} (\frac{s}{w})^N$ gives non-zero contribution after integration over $w$. Application of formulas (\ref{eq:PfromU}--\ref{eq:SfromV}) completes the proof.
\qed

\subsection{Saddle point analysis} \label{sec:saddle_point}

Let us recall the relation $\cP_N(z,t) = (1-\tau)^{-1} \cP_N^{\tau}(z,q=\frac{t}{1-\tau})$, which means that it is sufficient to study the asymptotics of $\cP_N^{\tau}(z,q)$, where $q=\frac{N^{1/3}t}{b^2}$.

\begin{proposition} \label{th:QN}
Let $z=\sqrt{N}(1+\tau) + \zeta N^{-1/6}$, $\tau = 1-b^2 N^{-1/3}$, $q=\frac{N^{1/3}t}{b^2}$ with $b\in \mathbb{R}$ and $Q_N(z,q,\tau)$ be given by formula \eqref{eq:QN}. Then
\begin{equation}
\lim_{N\to \infty} N^{-1/6}e^{-\frac{z^2}{1+\tau}} Q_N(z,q,\tau) = \frac{4b^2}{t} \sqrt{2\pi}\left (T_0(\zeta) + \frac{b^2T_1(\zeta)}{t} + \frac{b^4T_2(\zeta)}{t^2} + \frac{b^6 T_3(\zeta)}{t^3}\right),
\end{equation}
where
\begin{eqnarray}
T_3(\zeta) & = &  \int_{\zeta}^{\infty} \left[ \Ai_b'^2(p) -  \Ai_b(p)\Ai_b''(p)\right] dp,
\\
T_2(\zeta) & = & b^2 T_3 + \int_{\zeta}^{\infty} \Ai_b^2(p) dp,
\\
T_1(\zeta) & = &  - \zeta T_3 + \frac{1}{2} \Ai_b^2(\zeta)  + b^2 \int_{\zeta}^{\infty} \Ai_b^2(p) dp,
\\
T_0(\zeta) & = & - T_3 +\frac{1}{2}b^2 \Ai_b^2(\zeta) - \frac{1}{2} \Ai_b(\zeta)\Ai_b'(\zeta)  - \zeta \int_{\zeta}^{\infty} \Ai_b^2(p)dp.
\end{eqnarray}
\end{proposition}

\textit{Proof.} To make the integral representation in Lemma~\ref{th:IntegralRepresentation} amenable for the saddle point analysis, we rescale the integration variables $(s,w)\to \sqrt{N}(s,w)$. Then, the dominant term in the integrand reads $\exp(N (f(s) - f(w))$, with
$f(s) = \frac{s^2}{2} - 2s + \ln s$. The function $f$ has one (doubly degenerate) stationary point at $s^*=1$.  At the stationary point $f''(s^*)=0$ and $f'''(s^*)= 2$, thus $f$ should be expanded up to the 3rd order in Taylor expansion and the stationary point should be probed on the scale $N^{-1/3}$.
 Close to the saddle point, we parameterize the integrals as $s=1+uN^{-1/3}$ and $w=1+vN^{-1/3}$, which leads to $f(s) = f(s^*) +  u^3/3 +b^2u^2/2- u\zeta$. 
 
 In order to collect the dominant contribution to the integral from the stationary point, the contours $C_{\varepsilon}$ and $\Gamma_{\delta}$ are deformed as presented in Fig~\ref{fig:contours}b. The stationary point is approached by the deformed contour $\Gamma_{\delta}$ at an angle $-\frac{\pi}{3}$ and is departed at an angle $+\frac{\pi}{3}$. Upon deforming $C_{\varepsilon}$ the stationary point is approached at an angle $-\frac{2\pi }{3}$ and departed at an angle $+\frac{2\pi }{3}$. The contours do not touch the stationary point and hence do not intersect. Since the dominant contribution to the integrals comes from vicinity of the stationary point (remaining contributions are exponentially suppressed at large $N$), the incoming and departing parts of the contours can be extended by straight lines going to infinity at prescribed angles, see Fig~\ref{fig:contours}c. We denote the contours schematically as $\cont$ and $\contSecond$.

 Before going into the saddle point calculations, let us first argue that the integrands are exponentially suppressed on the full length of the deformed contours except close to the stationary point.  With the use of a simple inequality 
\begin{equation}
\left|\int \exp(-N f(w)) dw \right| \leq \int \left|\exp(-N f(w))\right| d|w| = \int \exp(-N \textup{Re}f(w)) d|w|
\end{equation}
it remains to show that along the contour $\textup{Re} f(w)>0$ and  $\textup{Re} f(s) < 0$. In fact, the dominant term is $e^{N(f(s)-f(w))}$, hence it is convenient to consider a shifted function $\tilde{f} = f - \frac{3}{2}$. Fig~\ref{fig:contours}b, presents the region where $\textup{Re} \tilde{f}(w)>0$ and the  $w$-integral is exponentially suppressed. Analogously, the $s$-integration contour passes through the complement of that region, hence the dominant contribution of the double integral indeed comes from the vicinity of the stationary point.

% 
%  which we denote as $\cont$. Due to the fact that $f(w)$ appears in the exponent with the minus sign, the contour $C_{\varepsilon}$ is deformed into a ray incoming at angle $-\frac{2\pi i}{3}$ and outgoing at $+\frac{2\pi i}{3}$, denoted as $\contSecond$. The contours do not reach the stationary point and do not intersect, see Fig.~\ref{fig:contours}b.

Let us now first apply the saddle point calculation for the asymptotics of $\tP_N$.
 After neglecting the contribution far away from the stationary point and extending the rays to infinity, we obtain
\begin{equation}
e^{-\frac{z^2}{1+\tau}} \tP_N(z) =  -N^{5/6}\frac{\sqrt{2\pi}}{(2\pi i )^2} \int\limits_{\contSecond} dv \int\limits_{\cont} du\, e^{\frac{u^3}{3} + \frac{b^2u^2}{2} - u\zeta}e^{-\frac{v^3}{3} + \frac{b^2v^2}{2} + v\zeta} \frac{u+v}{u-v} + \mathcal{O}(N^{1/2}).
\end{equation}
We observe that the exponents differ by a sign in the cubic terms. Therefore, it is convenient to change $v\to -v$ in the integral to symmetrize them. This transformation maps the contour $\contSecond$ into $\contMinusFirst$, which is the same as $\cont$, but with opposite orientation. The orientation can be easily changed, absorbing the minus sign resulting from the change of variables. % The above considerations can be summarized into reparameterization $s=1 + u N^{-1/3}$, $w=1 - v N^{-1/3}$ around the stationary point, where the integrals over $u$ and $v$ are performed over the contour $\cont$. 
This leads to
\begin{equation}
e^{-\frac{z^2}{1+\tau}} \tP_N(z) =  -N^{5/6}\frac{\sqrt{2\pi}}{(2\pi i )^2} \int\limits_{\cont} dv \int\limits_{\cont} du\, e^{\frac{u^3}{3} + \frac{b^2u^2}{2} - u\zeta}e^{\frac{v^3}{3} + \frac{b^2v^2}{2} - v\zeta} \frac{u-v}{u+v} + \mathcal{O}(N^{1/2}).
\end{equation}
It is now easy to see that the integrand is antisymmetric after exchanging $u\leftrightarrow v$, while the contours are the same, thus the integral evaluates to 0. Therefore, since $\tP_N$, and analogously $\tR_N$ with $\tS_N$, vanish in the leading order, a more careful saddle point analysis is required that goes into next orders in asymptotic expansion, see e.g.~\cite{asymptotic}.

To this end, we use the integrals representations in Lemma~\ref{th:IntegralRepresentation}. Following the reparameterizations around the saddle point, $s = \sqrt{N} + N^{1/6}u$, $w = \sqrt{N} - N^{1/6}v$, $z=\sqrt{N}(1+\tau) + \zeta N^{-1/6}$, $\tau = 1-b^2 N^{-1/3}$, the exponent is transformed into
\begin{equation}
- \frac{z^2}{1+\tau} + \frac{(s-z)^2}{2\tau} -\frac{\tau w^2}{2} + wz + N\ln\left(\frac{s}{w}\right) = \frac{u^3}{3} + \frac{b^2u^2}{2} - u\zeta  + \frac{v^3}{3} + \frac{b^2v^2}{2} - v\zeta + \Phi_{res},
\end{equation}
where the residual terms are subleading in $N$ and read
\begin{multline}
\Phi_{res} = \frac{v^4-u^4+2b^4u^2-4b^2u\zeta}{4N^{1/3}} + \frac{1}{N^{2/3}}\left(\frac{v^5+u^5}{5}+\frac{b^6u^2}{2}-b^4u\zeta + \frac{b^2\zeta^2}{4}\right) 
\\
+ \frac{4v^6-4u^6 + 12 b^8 u^2 - 24 b^6u\zeta + 9b^4\zeta^2}{24N} + \frac{1}{N^{4/3}}\left(\frac{v^7+u^7}{7} + \frac{b^{10} u^2}{2} - b^8 u\zeta + \frac{7b^6\zeta^2}{16}\right) + \mathcal{O}\left(N^{-5/3}\right).
\end{multline}
As it turns out later, we needed to expand $\Phi_{res}$ up to the order $N^{-4/3}$, since first three orders of the expansion around saddle point vanish and the first non-vanishing contribution involves the terms of order $N^{-4/3}$. The term $e^{\Phi_{res}}$ is systematically expanded into a Taylor series, together with the preexponential terms, where we also reparameterize $q= N^{1/3}t / b^2$. The exact tedious calculations are systematically performed in Mathematica, while here we outline the technique and key computational steps. The next terms in the expansion for $Q_N$ reads
\begin{multline}
e^{-\frac{z^2}{1+\tau}} Q_N(z,q,\tau) = N^{5/6}\frac{2b^2\sqrt{2\pi}}{(2\pi i )^2t} \int\limits_{\cont} dv \int\limits_{\cont} du\,e^{\frac{u^3}{3} + \frac{b^2u^2}{2} - u\zeta}e^{\frac{v^3}{3} + \frac{b^2v^2}{2} - v\zeta} \frac{2 + b^2 (u-v)^2-u^2 v-u v^2+u^3+v^3}{u+v}
\\
+ \mathcal{O}(N^{1/2}). \label{eq:first_order}
\end{multline}
The representation $a^{-1} = \int_0^{\infty} e^{-ap}dp$ for $\textup{Re}(a) >0$ together with~\eqref{eq:AiryDeformed} allow us to rewrite 
\begin{equation}
\frac{1}{(2\pi i )^2} \int\limits_{\cont} dv \int\limits_{\cont} du\, e^{\frac{u^3}{3} + \frac{b^2u^2}{2} - u\zeta}e^{\frac{v^3}{3} + \frac{b^2v^2}{2} - v\zeta} \frac{u^kv^l}{u+v} = (-1)^{k+l}\int_{0}^{\infty} \Ai_b^{(k)}(p+\zeta)\Ai_b^{(l)}(p+\zeta) dp,
\end{equation}
where $\Ai_b^{(k)}$ denotes the $k$-th derivative. The integral in \eqref{eq:first_order} is then evaluated into
\begin{multline}
N^{5/6}\frac{4b^2 \sqrt{2\pi}}{t} \int_{0}^{\infty} \left( \Ai_b^2(p+\zeta) + b^2 \Ai_b(p+\zeta)\Ai''_b(p+\zeta) - \Ai_b(p+\zeta)\Ai_b^{(3)}(p+\zeta) \right.
\\ 
\left. - b^2 \Ai'^2_b(p+\zeta) + \Ai'_b(p+\zeta)\Ai''_b(p+\zeta) dp \label{eq:first_order_reduced}
\right) .
\end{multline}
Higher order derivatives can be expressed by the lower order ones with the use of the relation
\begin{equation}
\Ai_b^{(k)}(p+\zeta) = b^2 \Ai_b^{(k-1)}(p+\zeta) + (p+\zeta)\Ai_b^{(k-2)}(p+\zeta) + (k-2)\Ai_b^{(k-3)}(p+\zeta), \label{eq:AiryDerivatives}
\end{equation}
which is derived from the second order equation $\Ai''_b(p) = b^2\Ai'_b(p)+p\Ai_b(p)$. After systematic application of \eqref{eq:AiryDerivatives}, the expression in \eqref{eq:first_order_reduced} is evaluated to 0. With the similar reasoning applied, all terms of order $N^{1/2}$ can be shown to cancel out as well.

At order $N^{1/6}$ the systematic technique needs to be enhanced due to the contribution from the fourth term in \eqref{eq:QN} (it was subleading at higher orders). Specifically, the term $\frac{1}{(s-w)^2}$ in \eqref{eq:Srepr} leads to the terms $\frac{u^k v^l}{(u+v)^2}$, which are represented with the use of $a^{-2} = \int_0^{\infty} p e^{-ap}dp$. Combined with \eqref{eq:AiryDerivatives}, this introduces terms 
\begin{equation}
A_k = \int_{0}^{\infty} p^k \Ai'^2_b(p+\zeta) dp,  \qquad B_k = \int_{0}^{\infty} p^k  \Ai^2_b(p+\zeta)dp, \qquad C_k = \int_{0}^{\infty} p^k \Ai'_b(p+\zeta) \Ai_b(p+\zeta) dp.
\end{equation} 
The power of $p$ in the integral is systematically reduced with the use of the following recurrence relations:
\begin{equation}
\begin{array}{r c l r}
2b^2 A_k & = & -k A_{k-1} - 2\zeta B_k + (k+1)C_k & k \geq 1,
\\
B_k & = & -A_{k-1} - \zeta B_{k-1} - b^2 C_{k-1} - (k-1)C_{k-2} & k \geq 2,
\\
C_k & = & -\frac{k}{2} B_{k-1} & k\geq 1.
\end{array}
\end{equation}
Finally, we use $B_1=-b^2C_0 - \zeta B_0 + \int_0^{\infty} \Ai_b(p+\zeta)\Ai_b''(p+\zeta) dp$ and shift the integration variable $p\to p-\zeta$.
\qed

\vspace{1cm}
\textit{Proof of Theorem~\ref{th:Main}.} Having worked out the asymptotics of $Q_N$ in Proposition~\ref{th:QN}, the remaining step is to take the limit of the terms in the exponent after the scaling $z=\sqrt{N}(1+\tau) + \zeta N^{-1/6}$, $\tau = 1-b^2 N^{-1/3}$, $q=\frac{N^{1/3}t}{b^2}$. It reads
\begin{equation}
-\frac{z^2}{2(1+\tau)}\left(\frac{q}{q+1}-1\right) + \frac{N}{2}\ln\left(\frac{q}{q+1+\tau}\right) = \frac{b^2\zeta}{t} -\frac{b^6}{2t^2} - \frac{b^6}{3t^3} + \mathcal{O}(N^{-1/3}).
\end{equation}
\qed

\subsection{From edge to bulk weak non-normality}
\label{sec:BulkWeak}
\textit{Proof of Corollary~\ref{th:BulkWeak}.}
From the left tail asymptotics of the Airy function $\Ai(-x) \sim \frac{1}{\sqrt{\pi} x^{1/4}} \sin(\frac{2}{3}x^{3/2} + \frac{\pi}{4})$ and its derivative $\Ai'(-x) \sim - \frac{x^{1/4}}{\sqrt{\pi}} \cos( \frac{2}{3} x^{3/2} + \frac{\pi}{4})$, using $b^2=a^2 / 2\nu$, we obtain
\begin{eqnarray}
\Ai_b(-\nu^2 w) & \sim & \frac{1}{w^{1/4} \sqrt{\pi \nu}} e^{-\frac{a^2w}{4}} \sin\left(\frac{2}{3} (\nu^2 w)^{3/2} + \frac{\pi}{4}\right), \label{eq:AiryLeft}
\\
\Ai'_b(-\nu^2 w) & \sim & \frac{\sqrt{\nu}}{w^{1/4} \sqrt{\pi}} e^{-\frac{a^2w}{4}} \cos\left(\frac{2}{3} (\nu^2 w)^{3/2} + \frac{\pi}{4}\right).
\end{eqnarray}

To calculate the asymptotics of integrals appearing in expressions \eqref{eq:T3} - \eqref{eq:T0}, we split the integral as 
\begin{equation}
\int_{\zeta}^{\infty} \Ai_b^2(p)dp = \int_0^{\infty} \Ai_b^2(p)dp - \int_0^{\zeta} \Ai_b^2(p)dp.
\end{equation}
 The first integral is $\mathcal{O}(1)$, while the second one yields the dominant contribution
\begin{equation}
 - \int_0^{\zeta} \Ai_b^2(p)dp = \nu^2 w \int_0^{1} \Ai_b^2(-\nu^2 wq) dq \sim \frac{\nu\sqrt{w}}{\pi} \int_0^{1} \frac{dq}{\sqrt{q}} e^{-\frac{a^2 w q}{2}} \sin^2 \left(\frac{2}{3} (\nu^2 wq)^{3/2} + \frac{\pi}{4}\right).
\end{equation}
In the first equality we changed variables $p=\zeta q$, while in the second step we used \eqref{eq:AiryLeft}. With the use of $\sin^2x = \frac{1}{2} - \frac{1}{2}\cos 2x$, we split the integral into a slowly varying part and rapidly oscillating integrand. In the large $\nu$ limit the oscillatory part evaluates to 0, while in the slowly varying part we change the integration variable $q = s^2$, obtaining
\begin{equation}
\int_{\zeta}^{\infty} \Ai_b^2(p)dp \sim \frac{\nu \sqrt{w}}{\pi} \int_0^{1} e^{-a^2 s^2 w/2} ds.
\end{equation}

To calculate the asymptotics of $T_3$, we use the differential equation $\Ai''_b(p) = b^2 \Ai'_b(p) + p \Ai_{b}(p)$ and apply a similar reasoning as described above to get
\begin{equation}
T_3(\zeta) \sim \frac{2\nu^3 w^{3/2}}{\pi} \int_{0}^{1} s^2 e^{-\frac{a^2 s^2w}{2}} ds.
\end{equation}

A direct inspection shows that dominant terms in \eqref{eq:Main} are $T_0$ and $b^2 T_1/t$, while higher order terms in $b$ are subleading. Simple calculations lead to
\begin{equation}
\cP^{w.e.}(\zeta,t) \sim \frac{A\nu\sqrt{w}}{2\pi t^2} \int_0^{1} \left(1 + \left(\frac{A}{t} -2\right)s^2\right)e^{-As^2/2}ds, 
\end{equation}
where we used $A = a^2 w$. Integration by parts brings the above formula to \eqref{eq:BulkWeak}.

\qed

\subsection{Strong non-normality limit} \label{sec:strong_nonnorm}
\textit{Proof of Corollary~\ref{th:StrongNonnormality}.} From the right tail asymptotics of the Airy function $\Ai(x)\sim  \frac{\exp(-\frac{2}{3}x^{3/2})}{2\sqrt{\pi} x^{1/4}}$  and its derivative $\Ai'(x) \sim -\frac{x^{1/4}}{2\sqrt{\pi}}\exp(-\frac{2}{3}x^{3/2})$ followed by the expansion to the second order
\begin{equation}
\left(b\delta \sqrt{2} + \frac{b^4}{4}\right)^{3/2} = \frac{b^6}{8}\left(1 + \frac{6\sqrt{2}\delta}{b^3} + \frac{12 \delta^2}{b^6} \right)+ \mathcal{O}\left(b^{-3}\right)
\end{equation}
 we get the following
\begin{equation}
\Ai_b(b\delta\sqrt{2}) = \frac{1}{b\sqrt{2\pi}} e^{-\delta^2} + \mathcal{O}\left(\frac{1}{b^2}\right),\qquad \qquad
\Ai'_b(b\delta\sqrt{2}) = -\frac{\delta}{b^2\sqrt{\pi}}e^{-\delta^2} + \mathcal{O}\left(\frac{1}{b^3}\right). \label{eq:AiryAsymptotics}
\end{equation}
Alternatively, this asymptotics can be obtained from the integral representation \eqref{eq:AiryDeformed} with the use of the saddle point method. The above result in particular means that $T_3(b\delta\sqrt{2})=\mathcal{O}(b^{-1})$ and since
\begin{equation}
\int_{b\delta \sqrt{2}} ^{\infty} \Ai_b^2(p)dp = b\sqrt{2} \int_{\delta}^{\infty} \Ai_b^2(bp\sqrt{2})dp = \frac{1}{2b\pi\sqrt{2}} \int_{2\delta}^{\infty} e^{-\frac{p^2}{2}} dp + \mathcal{O}\left(\frac{1}{b^2}\right), \label{eq:PropertyAi}
\end{equation}
we also have $T_2(b\delta\sqrt{2}) = \mathcal{O}(b)$. The exact asymptotics of $T_3$ and $T_2$ are not needed, since these terms are suppressed by $t^3$ and $t^2$ in the denominator in equation~\eqref{eq:Main}. Recall that the transition from weak to strong non-normality requires parameterization $t=\sqrt{2}b^3\sigma$ with $\sigma = \mathcal{O}(1)$. In the second equality in \eqref{eq:PropertyAi} we used the asymptotics \eqref{eq:AiryAsymptotics} under the integral, which can be justified by the dominated convergence theorem.  This was followed by rescaling of the integration variable $p\to p/2$. While formally justifying the assumptions of the dominated convergence theorem is a rather technical task, the Gaussian asymptotics in \eqref{eq:PropertyAi} provides hints for the applicability of the theorem. For example, a function $g(p) = \exp(-|p|)$ is a good candidate for the dominating function of $f_b(p) = 2b^2 \Ai_b^2(bp\sqrt{2})$.  Using \eqref{eq:PropertyAi}, we calculate the asymptotics of the remaining terms
\begin{eqnarray}
T_1(b\delta\sqrt{2}) & = & \frac{b}{2\pi\sqrt{2}}\int_{2\delta}^{\infty} e^{-p^2/2}dp + \mathcal{O}(1),
\\
T_0(b\delta\sqrt{2}) & = & \frac{1}{4\pi}e^{-2\delta^2} - \frac{\delta}{2\pi} \int_{2\delta}^{\infty} e^{-p^2/2}dp + \mathcal{O}\left(\frac{1}{b}\right).
\end{eqnarray}
This result combined with taking  remaining straightforward limits completes the proof.
\qed

\subsection{Integrating out the self-overlap} \label{sec:integrating_out}

We start with two properties satisfied by the deformed Airy functions that are useful in manipulations of the formulas. They are straightforward to show, so we present them without a proof.

\begin{lemma} \label{th:AiryProperties}
Let $\Ai_b(\zeta) = \exp(\frac{1}{2}b^2\zeta + \frac{1}{12}b^6)\Ai(\zeta+\frac{b^2}{4})$ and $\Bi_b(\zeta) = \exp(\frac{1}{2}b^2\zeta + \frac{1}{12}b^6)\Bi(\zeta+\frac{b^2}{4})$, where $\Ai$ and $\Bi$ are the Airy functions. Let $F_b$ and $G_b$ be any combination of $\Ai_b$ and $\Bi_b$. They satisfy the relations
\begin{eqnarray}
F_{ib}(\zeta) G_{b}(\zeta) & = &  F_{b}(\zeta)G_{ib}(\zeta)  = F_0\left(\zeta+\frac{b^4}{4}\right)G_0\left(\zeta+\frac{b^4}{4}\right), \label{eq:AiryProp1}
\\
F'_{ib}(\zeta)G_b(\zeta) & = & -\frac{b^2}{2} F_{ib}(\zeta) G_b(\zeta) + F_0'\left(\zeta+\frac{b^4}{4}\right)G_0\left(\zeta+\frac{b^4}{4}\right). \label{eq:AiryProp2}
\end{eqnarray}
\end{lemma}

\vspace{0.3cm}

We are now ready to calculate the integral giving rise to the deformed Scorer's function.

\noindent \textit{Proof of Lemma~\ref{th:Scorers}.} The exact form of the deformed Scorer's function can be found by the variation of parameters method, since $\Hi_b$ satisfies the inhomogeneous second order differential equation
\begin{equation}
\Hi_b''(\zeta) + b^2 \Hi'_b(\zeta) - \zeta \Hi_b(\zeta) = \frac{1}{\pi}, \label{eq:ScorerDiffEq}
\end{equation}
which follows from $1 = \int_0^{\infty}(u^2 + b^2u - \zeta )\exp(\zeta u - \frac{b^2u^2}{2} - \frac{u^3}{3}) du $. Instead of proving that \eqref{eq:Scorer} matches the asymptotics, we observe that $\Hi_b$ satisfies the (backward) diffusion equation in the variable $\eta=\frac{b^2}{2}$ playing the role of time. That is, the function $h(\zeta,\eta) = \pi^{-1}\int_0^{\infty}\exp(\zeta u - \eta u^2 - \frac{u^3}{3}) du$ satisfies equation $\partial_{\eta} h = -\partial_{\zeta\zeta}h$ with the initial condition given by the Scorer's function
\begin{equation}
h(\zeta,0)= \Bi(\zeta)\int_{-\infty}^{\zeta} \Ai(t) dt -\Ai(\zeta)\int_{-\infty}^{\zeta} \Bi(t) dt.
\end{equation}
Trivially, \eqref{eq:Scorer} reduces to the above as $b\to 0$. Straightforward verification that \eqref{eq:Scorer} satisfies $\partial_{\eta} h = -\partial_{\zeta\zeta}h$ uses Lemma~\ref{th:AiryProperties} and the fact that $\Ai_b$ and $\Bi_b$ themselves satisfy diffusion equation, see Remark~\ref{remark}.

\qed

\vspace{0.3cm}

\textit{Proof of Corollary~\ref{th:Integration}.}
To integrate out the $t$ variable in \eqref{eq:Main}, we first change the integration variable to $u=\frac{b^2}{t}$ and rewrite the resulting integral as
\begin{multline}
\int_0^{\infty} \cP^{w.e.}(\zeta,t)dt = \int_0^{\infty} e^{-\frac{u^3}{3}-\frac{b^2u^2}{2} + \zeta u} \left[ (u^3+b^2u^2-\zeta u-1)T_3(\zeta) + (u^2+b^2u-\zeta) \int_{\zeta}^{\infty} \Ai^2_b(p)dp \right.
\\
\left.
+\frac{1}{2} u \Ai^2_b(\zeta) + \frac{1}{2}b^2\Ai_b^2(\zeta) - \frac{1}{2}\Ai_b(\zeta)\Ai'_b(\zeta)
\right]du.
\end{multline}
The integrals can be expressed in terms of $\Hi_b$ and its derivatives. However, we notice that the first parenthesis in the square bracket integrates out to 0, because it corresponds to the third order differential equation satisfied by $\Hi_b$, which can be obtained by differentiating both sides of~\eqref{eq:ScorerDiffEq}. Analogously, the second parenthesis integrates to 1, by the virtue of~\eqref{eq:ScorerDiffEq}. The fact that 
\begin{displaymath}
\pi\Ai_b(\zeta) \Hi'_b(\zeta) + b^2\pi\Ai_b(\zeta)\Hi_b(\zeta) - \pi\Ai'_b(\zeta)\Hi_b(\zeta) = \int_{-\infty}^{\zeta}\Ai_b(t) dt
\end{displaymath}
can be verified by explicit calculations that involve the exact form of the deformed Scorer's function (see Lemma~\ref{th:Scorers}) and simple manipulations using Lemma~\ref{th:AiryProperties}, followed by the use of the Wronskian $\Ai(x)\Bi'(x) - \Ai'(x) \Bi(x)=\pi^{-1}$. The final observation 
\begin{equation}
\int_{-\infty}^{\infty}\Ai_b(t)dt = \int_{-\infty}^{\infty}\Ai(t)dt = 1
\end{equation}
completes the proof. The first equality follows from the fact that $\Ai_b$ is a solution to the diffusion equation (see Remark~\ref{remark}) and diffusion preserves probability, while the second equality is a known result.
\qed 

\section*{Acknowledgements}
The author is grateful to T. R. Würfel, Y. V. Fyodorov, A. Miroszewski and M. A. Nowak for discussions.

\end{document}